\documentclass{svproc}
\usepackage{graphics}
\usepackage{graphicx}

\usepackage{bbm}
\usepackage{bm}

\usepackage{amsfonts,cite}

\usepackage{hyperref}

\usepackage{url}

\def\={\ =\ }
\def\alg{{\mathcal A}}
\newcommand{\Tr}[1]{\:{\rm Tr}#1}
\def\ii{{\,{\rm i}\,}}
\def\e{{\,\rm e}\,}
\newcommand{\fru}{{\mathfrak u}}

\newcommand{\bbr}{{\mathbbm R}}
\newcommand{\bbc}{{\mathbbm C}}
\def\ds{\stackrel{\star}{,}}
\newcommand{\End}{{\rm End}}
\newcommand{\frg}{{\mathfrak g}}
\def\dd{{\rm d}}

\begin{document}
\mainmatter              
\title{Gravity versus Noncommutative Gauge Theory: A Double Copy Perspective}
\titlerunning{Gravity vs Noncommutative Gauge Theory}  
%
\author{Richard J. Szabo}
\authorrunning{Richard J. Szabo} 
%
\tocauthor{Richard J. Szabo}
\institute{Department of Mathematics, Heriot-Watt University\\  Maxwell Institute for Mathematical Sciences, Edinburgh\\
\email{R.J.Szabo@hw.ac.uk}}

\maketitle              

\begin{abstract}
We discuss how Moyal deformations of gauge theories, which arise naturally from open string theory, fit into the paradigm of colour-kinematics duality and the double copy of gauge theory to gravity. Along the way we encounter novel noncommutative scalar field theories with rigid colour symmetry that have no interacting commutative counterparts. These scalar theories offer  new perspectives on old ideas that rank one noncommutative gauge theories are gravitational theories. This is rendered explicit in four dimensions where they and their double copy images yield deformations of integrable theories describing the self-dual sectors of Yang-Mills theory and gravity.
\end{abstract}
\section{Introduction}

It is an endearing pleasure for me to write this contribution on the occasion of Gordon Semenoff's 70th birthday. Gordon was my MSc and then PhD supervisor at the University of British Columbia from 1990--1995. From him I learned how to become a careful and independent researcher. 
The present article describes interactions between quantum field theory, string theory, gravity and mathematical physics, which are themes that have permeated Gordon's research interests.

The basic premise of this contribution is the double copy of noncommutative Yang-Mills theory, as it appears in low-energy effective field theories on D-branes in constant $B$-fields, to gravity, which we describe in \S\ref{sec:doublecopyNCYM} below. It is based on the lengthy recent paper~\cite{Szabo:2023cmv}, which is engulfed in the modern language of the homotopy algebraic approach to quantum field theory. In the remaining sections we give a concise summary of some of the main points of that work, in a manner which we hope is more palatable to a broader audience. We skip most technical details, while referring to~\cite{Szabo:2023cmv} for all mathematical constructions as well as more thorough and complete citations to the relevant literature. 

\section{Double Copying Noncommutative Yang-Mills Theory}
\label{sec:doublecopyNCYM}

\paragraph{\bf KLT Double Copy Relations.} 

The double copy of gauge theory to gravity has its origins in string theory, going back to the mid-1980s with the famous relations between tree-level scattering amplitudes of closed and open strings discovered by Kawai, Lewellen and Tye (KLT)~\cite{Kawai:1985xq}. These relations reflect the topological property that a sphere, representing a closed string worldsheet, can be realised by gluing together two discs, representing open string worldsheets:
$$
\includegraphics[width=7cm]{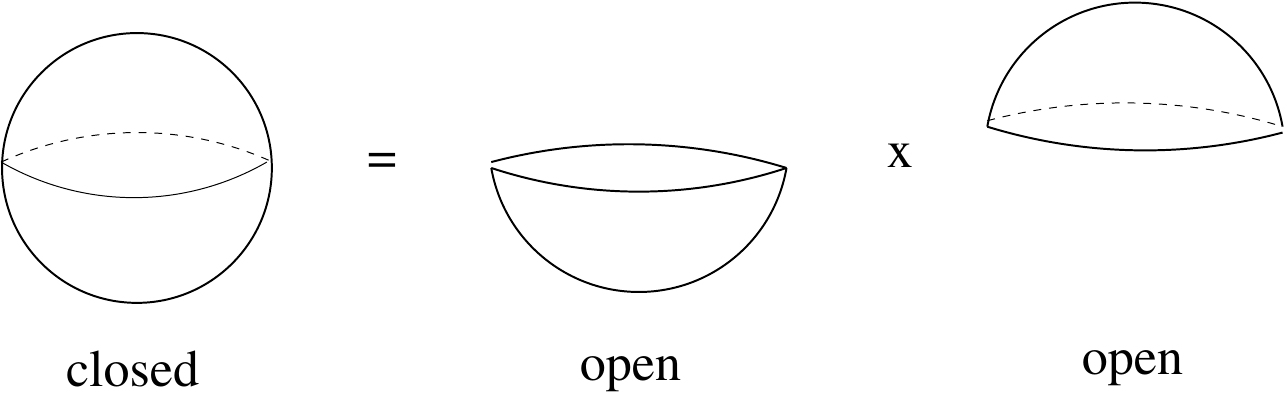}
$$

In the field theory limit $\alpha'\longrightarrow0$, the left-hand side represents a tree-level amplitude for graviton scattering while the right-hand side consists of amplitudes of gluons. This results in the famous double copy slogan:
 \begin{equation} \label{eq:DCslogan}
 \textrm{Gravity} \ = \ (\textrm{Gauge Theory})^2 \ .
 \end{equation}
At a more quantitative level, the KLT double copy relations between $n$-point amplitudes read
 \begin{equation} \label{eq:KLT}
 M_n^{L\otimes R} \= \sum_{w,w'}\,A_n^L(w) \ \boldsymbol S_n(w|w') \ A_n^R(w') \ ,
 \end{equation}
 where the sums run over orderings of $n$ external gluons, and $A_n$ are colour-stripped amplitudes appearing in the full gluon amplitudes 
\begin{equation} 
 \alg_n\= \sum_{S_n/\mathbbm{Z}_n} \, \Tr\,(T^{a_1}\cdots T^{a_n}) \ A_n(1,\dots,n) 
\end{equation} 
 expressed as a sum over cyclic orderings of external particles. 
 
The quantity $\boldsymbol S_n(w|w') $ which glues left and right gauge theory amplitudes together is called the \emph{KLT momentum kernel}. It was realised almost 30 years after the advent of the KLT relations that they can be expressed as the inverses 
 \begin{equation} \label{eq:KLTmomker}
\boldsymbol S_n(w|w') \=\boldsymbol m_n(w|w') ^{-1}
 \end{equation}
of bi-coloured scalar amplitudes $\boldsymbol m_n(1,\dots,n|1,\dots,n)$ in a field theory known as the biadjoint scalar theory~\cite{Cachazo:2013iea}, which we shall discuss in more detail later on.

\paragraph{\bf Noncommutative Field Theory.}

In the late 1990s it was realised that, on $\bbr^d$ with a constant NS--NS $B$-field and flat background metric $g$, the low-energy interactions of open strings ending on stacks of $N$ D-branes are described by a non-local deformation of $U(N)$ Yang-Mills theory to a \emph{noncommutative gauge theory}~\cite{Douglas:1997fm,Ardalan:1998ce,Chu:1998qz,Schomerus:1999ug,Seiberg:1999vs,Douglas:2001ba,Szabo:2001kg}. Open strings see an effective metric $G$ and a bivector $\theta$ related to the closed string background $(g,B)$, such that $G=g$ and $\theta=0$ when $B=0$.

Colour-stripped open string amplitudes then have a simple $\theta$-dependence given by  phase factors
\begin{equation} \label{eq:AnGtheta}
A_n(1,\dots,n)_{G,\theta} \= \e^{-\ii\sum_{i<j}\,p_i\times p_j} \ A_n(1,\dots,n)_{G,\theta=0}  \ , 
\end{equation}
where $ k\times q\=\mbox{$\frac12$}\,
k_\mu\,\theta^{\mu\nu}\,q_\nu$ and $p_i$ are the external momenta. The right-hand side involves the open string amplitude in the absence of the $B$-field but computed with the open string metric $G$. 
In the Seiberg-Witten scaling limit where $\alpha'\longrightarrow0$  with $g\sim(\alpha')^2\longrightarrow0$, these phase factors are captured by replacing the commutative pointwise product $\mu(\phi_1\otimes\phi_2):=\phi_1\,\phi_2$ of fields $\phi_1,\phi_2\in \Omega^0(\bbr^d,{\mathfrak u}(N))$ with their associative Moyal product defined by
\begin{equation} \label{eq:Moyalprod}
\phi_1\star \phi_2 \= \mu \circ \exp\big(\mbox{$\frac{\ii}2$}\,
\theta^{\mu\nu}\, \partial_\mu\otimes\partial_\nu\big)(\phi_1\otimes \phi_2)  \ .
\end{equation}

\paragraph{\bf KLT Relations with $\boldsymbol B$-Fields.}

The $B$-field modifications of the KLT relations are trivial: we simply rewrite the original KLT formula (\ref{eq:KLT}) by substituting (\ref{eq:AnGtheta}) to get
\begin{eqnarray}
 M_n^{L\otimes R} & \= &\sum_{w,w'}\,A_n^L(w)_{G,\theta} \ \boldsymbol S_n(w|w')_{G;\theta,\bar\theta} \ A_n^R(w')_{G,\bar\theta} \ ,
\end{eqnarray}
where the gravitational amplitudes on the left-hand side are now computed with the background metric $G$. The new momentum kernel is
\begin{eqnarray} \label{eq:KLTBfield}
\boldsymbol S_n(w|w')_{G;\theta,\bar\theta} & \= &\Theta_n(w)^{-1} \ \boldsymbol S_n(w|w')_{G;\theta=0,\bar\theta=0} \ \bar\Theta_n(w')^{-1} \ , 
\end{eqnarray}
where
\begin{equation} \label{eq:Thetan}
\quad \Theta_n(1,\dots,n)  \= \e^{-\ii\sum_{i<j}\,p_i\times p_j} \ .
\end{equation}
We have allowed for different bivectors $\theta$ and $\bar\theta$ in the left and right gauge theory amplitudes, which appear when distinct gauge fluxes on the worldvolumes of different stacks of D-branes are turned on.

So far this is a boring result: it simply modifies the slogan (\ref{eq:DCslogan}) to
\begin{equation}
\textrm{Gravity} \= (\textrm{Noncommutative Gauge Theory})^2 \ .
\end{equation}
That is, the double copy of noncommutative Yang-Mills theory is \emph{ordinary} general relativity, rather than some  deformation of gravity as one may have anticipated. However, this result raises several theoretical questions; for example:
\begin{itemize}
\item[] {\sf \underline{Question 1:}} \ {What is the scalar theory sourcing the relation $\boldsymbol S_n=\boldsymbol m_n^{-1}$ for the $B$-field modified KLT momentum kernel?}
\end{itemize}
The answers to this and other questions lead to new perspectives on the double copy relations, which require  careful considerations of non-locality, with interesting answers and applications. These are the topic of the remaining sections.

\section{Colour and Kinematics in Moyal-Deformed Theories}
\label{sec:CKduality}

\paragraph{\bf Colour-Kinematics Duality.}

Fastforwarding ahead 20 years from the KLT paper, Bern, Carrasco and Johansson (BCJ)~\cite{Bern:2008qj} posit a stripping of colour from kinematic factors in numerators of tree-level gauge theory amplitudes in the form
\begin{equation} \label{eq:BCJ}
\alg_n^{L/R} \ = \ \displaystyle\sum\limits_I \, \frac{c_I\,n_I^{L/R}}{D_I} \ ,
\end{equation}
where the sum runs over all planar trivalent Feynman graphs with $n$ external edges, $c_I$ are colour factors constructed from structure constants $f^{abc}$ of the gauge algebra $\frg$, $n_I$ depend only on purely kinematic data (momentum, spin, flavour, etc.), and $D_I$ are propagator factors associated to internal edges. 

The colour-stripping in (\ref{eq:BCJ}) depends on a choice of basis, which is dictated by demanding that it be compatible with \emph{colour-kinematics duality}: any relation of the schematic form $c_I+c_J+c_K+\cdots=0$ involving only colour factors, such as the Jacobi identity among $f^{abc}$ for the Lie algebra $\frg$, is mirrored by a relation $n_I+n_J+n_K+\cdots=0$ among kinematic factors. These are called generalized Jacobi identities. The duality predicts the existence of an infinite-dimensional kinematic algebra underlying scattering amplitudes, whose general structure is currently one of the main open problems in the double copy programme.

When the colour-stripping is compatible with colour-kinematics duality, then the double copy relations replace colour by kinematics \smash{$c_I\longmapsto n_I^{R/L}$}  in (\ref{eq:BCJ}) to give amplitudes of a theory with linearized diffeomorphism symmetry via two copies of gauge theory kinematic factors:
\begin{equation}
M_n^{L\otimes R} \ = \ \displaystyle\sum\limits_I \, \frac{n_I^{L}\,n_I^{R}}{D_I} \ .
\end{equation}
The BCJ relations ensure that the amplitudes are independent of the chosen basis for the decomposition (\ref{eq:BCJ}). 

This operation also allows one to replace kinematic factors \smash{$n_I^{L/R}\longmapsto \bar c_I$} in (\ref{eq:BCJ}) by colour factors constructed from structure constants $\bar f^{\bar a\bar b\bar c}$ of another gauge algebra $\bar\frg$, regarded as a kinematic Lie algebra since the resulting amplitudes trivially satisfy colour-kinematics duality. This gives the amplitudes of a  \emph{biadjoint scalar theory}, called the ``zeroth copy'' of the gauge theory because it acts as the identity theory in the space of copyable theories: the double copy of any field theory with the biadjoint scalar theory simply returns the original theory.

These considerations now bring us to our second question:
\begin{itemize}
\item[] {\sf \underline{Question 2:}} \ {What is the meaning of colour-kinematics duality in noncommutative gauge theories?}
\end{itemize}
To understand why this is a question, we review the well-known mixing between gauge and kinematic degrees of freedom in Moyal-deformed field theories.

\paragraph{\bf Colour-Kinematics Mixing.}

Consider the noncommutative associative algebra $\Omega^0(\bbr^d,{\mathfrak u}(N))$ with the Moyal product. The star-commutator
\begin{equation}
[\phi_1\ds\phi_2] \= \phi_1\star\phi_2 - \phi_2\star\phi_1
\end{equation}
makes it an infinite-dimensional Lie algebra, which we denote as ${\mathfrak u}_\star(N)$. As a vector space, 
$\Omega^0(\bbr^d,{\mathfrak u}(N)) \simeq {\mathfrak u}(N)\otimes \Omega^0(\bbr^d)$, but not as the Lie algebra  ${\mathfrak u}_\star(N)$  unless  $\theta  =  0$, because expanding the fields in a basis $\{T^a\}$ of $\fru(N)$ shows
\begin{equation}
[\phi_1\ds\phi_2] \= \mbox{$\frac12$}\,\big( f^{a b c}\,\{\phi_1^{ a}\ds\phi_2^{ b}\} + \ii d^{ a b c}\,[\phi_1^{ a}\ds\phi_2^{ b}]\big) \,  T^{ c} \ ,
\end{equation}
where $[T^a,T^b]=f^{abc}\,T^c$ and $\{T^a,T^b\}=\ii d^{abc}\,T^c$. This loss of factorization spoils standard colour-stripping and colour-kinematics duality.

\paragraph{\bf Kinematic Lie Algebras.}

On the other hand, this mixing provides a novel realisation of the kinematic algebras for some theories. The momentum space basis $e_k^a=\e^{\ii k\cdot x}\,T^a$ of $\Omega^0(\bbr^d,{\mathfrak u}(N))$ obeys the star-commutation relations
\begin{eqnarray}
\big[e_k^{ a} \ds e_p^{ b}\big] &\=& F^{ a b c}(k,p,k+p) \ e^{ c}_{k+p} \ ,
\end{eqnarray}
with the momentum dependent structure constants
\begin{eqnarray}
F^{ a b c}(k,p,k+p) &\=& f^{ a b c} \cos(k\times p) 
    +\ii d^{ a b c}\sin(k\times p) \ .
\end{eqnarray}

For $N=1$  this makes  $\Omega^0(\bbr^d)$ into the \emph{kinematic Lie algebra ${\mathfrak u}_\star(1)$} with
\begin{equation}
[e_k \ds e_p] = F(k,p,k+p) \ e_{k+p} \quad , \quad 
F(k,p,k+p) = 2 \ii\sin(k\times p) \ .
\end{equation}
In the semi-classical limit, i.e. to first order in the deformation bivector $\theta$, the star-commutator truncates to the Poisson bracket $\{\phi_1,\phi_2\}_\theta = \theta^{\mu\nu}\,\partial_\mu\phi_1\,\partial_\nu\phi_2$, and $\fru_\star(1)$ reduces to the Lie algebra ${\mathfrak{sdiff}}(\bbr^d)$  of \emph{symplectic diffeomorphisms}.
The full deformation ${\mathfrak u}_\star(1)$ of  ${\mathfrak{sdiff}}(\bbr^d)$  is
related to the infinite unitary algebra ${\mathfrak u}(\infty)$  as well as a certain $W_{1+\infty}$-algebra known as the symplecton algebra, see e.g.~\cite{Hoppe:1988gk,Pope:1989sr,Lizzi:2001nd,Chacon:2020fmr}. For a description of the $N>1$ kinematic Lie algebras $\fru_\star(N)$, see~\cite{Monteiro:2022xwq}.
    
\paragraph{\bf Twisted Colour-Stripping.}

The way to handle the mixing issue comes from a new perspective on colour-stripping~\cite{Szabo:2023cmv}, and ultimately a new notion of colour-kinematics duality, tailored to the momentum dependent structure constants of these inherently non-local field theories. It is based on the beautiful formalism of twisted tensor products of homotopy algebras developed by Borsten \emph{et al.}~\cite{Borsten:2021hua}, which has the additional power to provide off-shell formulations of colour-kinematics duality and the double copy relations. 

Skipping the formal definitions and details, the upshot is that the Lie algebra
\begin{equation}
\fru_\star(N) \= \fru(N)\otimes_\tau\Omega^0(\bbr^d)
\end{equation}
can be factorized as a \emph{twisted tensor product} of the Lie algebra $\fru(N)$ with the commutative algebra  $(\Omega^0(\bbr^d),\mu )$ of functions, where 
\begin{equation}
\tau:\fru(N)\otimes \fru(N)\longrightarrow \fru(N) \otimes\End\big(\Omega^0(\bbr^d)\big)\otimes\End\big(\Omega^0(\bbr^d)\big) 
\end{equation} 
is the \emph{twist datum} defined by
\begin{eqnarray}
\tau( T^{ a},  T^{ b}) & \= & [ T^{ a}, T^{ b}]\otimes\cos\big(\mbox{$\frac 12$}\,\theta^{\mu\nu} \, \partial_\mu \otimes \partial_\nu\big) \nonumber \\
& & \qquad + \, \ii\{ T^{ a}, T^{ b}\} \otimes \sin\big(\mbox{$\frac 12$}\,\theta^{\mu\nu} \, \partial_\mu \otimes \partial_\nu\big) \ .
\end{eqnarray}
Then in Sweedler notation
\begin{eqnarray}
[T^a\otimes\phi_1^a\ds T^b\otimes\phi_2^b] & \= & \tau^{(1)}(T^a,T^b) \nonumber \\
& & \qquad \otimes \ \mu\big(\tau^{(2)}(T^a,T^b)(\phi^a_1)\,,\,\tau^{(3)}(T^a,T^b)(\phi^b_2)\big) \ .
\end{eqnarray}

{This strips colour from kinematics, with each factor of the twisted tensor product a conventional algebraic object:} noncommutativity is now completely absorbed into the twist datum $\tau$.
For a given theory, this twisted factorization must be compatible with a form of colour-kinematics duality in order to be able to apply the double copy relation. This answers Question~2 above.

\paragraph{\bf Noncommutative Biadjoint Scalar Theory.}

Let us illustrate this operation by constructing a new non-local scalar field theory of relevance to our ensuing discussions. The \emph{biadjoint scalar theory}  is a massless cubic scalar field theory with rigid symmetry under a pair of quadratic Lie algebras $\frg$ and $\bar\frg$. Its fields 
\begin{equation}
\phi\=T^a\otimes\bar T^{\bar a}\otimes\phi^{a\bar a} \ \in \ \Omega^0(\bbr^d,\frg\otimes\bar\frg) \ \simeq \ \frg\otimes\big(\bar\frg\otimes\Omega^0(\bbr^d)\big)
\end{equation}
propagate according to the action functional
\begin{equation}
\displaystyle S_{{\rm bi}}[\phi]\=\int\,\dd^dx \ \frac12\,\phi^{a\bar a}\,\square\,\phi_{a\bar a}-\frac g{3!}\,f^{abc}\,\bar f^{\bar a \bar b\bar c}\, \phi^{a\bar a}\,\phi^{b\bar b}\,\phi^{c\bar c} \ .
\end{equation}

Deform the biadjoint scalar theory by replacing the Lie algebra $\bar\frg\otimes\Omega^0(\bbr^d)$ with the Lie algebra $\fru_\star(\bar N) \simeq \fru(\bar N)\otimes_{\bar\tau}\Omega^0(\bbr^d)$. The new action functional is
 \begin{equation}
  S_{{\rm bi}}^\star[\phi] \= \int\,\dd^dx \ \frac12\,\phi^{a\bar a}\star\square\,\phi_{a\bar a} - \frac g{3!} \,  f^{abc} \, \bar\lambda^{\bar a\bar b\bar c} \, \phi^{a\bar a} \star \phi^{b\bar b} \star \phi^{c\bar c} \ ,
  \end{equation}
where $\bar\lambda^{\bar a\bar b\bar c} := \bar f^{\bar a\bar b\bar c} + \ii\bar d^{\bar a\bar b\bar c}$. This theory has the standard scalar propagator, but now the 3-point Feynman vertex
\vspace{-3mm}
\unitlength=1.00mm
\linethickness{0.4pt}
$$
\hspace{-3cm}
\begin{picture}(100.00,25.00)
\thinlines
\put(50.00,12.00){\line(-1,1){6.00}}
\put(50.00,12.00){\line(-1,-1){6.00}}
\put(35.00,20.00){\makebox(0,0)[l]{{$k,a,\bar a$}}}
\put(36.00,4.00){\makebox(0,0)[l]{{$p,b,\bar b$}}}
\put(50.00,12.00){\line(1,0){8.00}}
\put(60.00,12.00){\makebox(0,0)[l]{$q,c,\bar c$}}
\put(68.00,12.00){\makebox(0,0)[l]{{$\=-\ii g\,f^{abc}\,\bar F^{\bar a\bar b\bar c}(k,p,q)$}}}
\end{picture}
\vspace{-3mm}
$$
comes from replacing the structure constants $\bar f^{\bar a\bar b\bar c}$ of $\bar\frg$ with the momentum dependent structure constants of $\bar F^{\bar a\bar b\bar c}(k,p,q)$ of $\fru_\star(\bar N)$ in the usual biadjoint scalar vertex.

\section{Noncommutative Chern-Simons Theory}
\label{sec:CS}

The simplest example in which to illustrate the new double copy prescription is the Moyal deformation of Chern-Simons gauge theory in three dimensions. The gluon fields are $\fru(N)$-valued 1-forms $A\in\Omega^1(\bbr^3,\fru(N))$ with action functional
\begin{equation}
S_{{\rm CS}}^\star[A] \= \int\, \mathrm{Tr}\, \Big(\frac12\,A\wedge_\star\dd A + \frac g{3!}\, A\wedge_\star [A \stackrel{\star}{,} A] \Big)  \ ,
\end{equation}
where the Moyal-deformed exterior product $\wedge_\star$ of forms is defined analogously to (\ref{eq:Moyalprod}) by replacing partial derivatives with Lie derivatives along the coordinate directions. The propagator is the same as in ordinary Chern-Simons theory, but now the 3-gluon Feynman vertex is
$$ \hspace{-3cm}
\includegraphics[width=2.5cm]{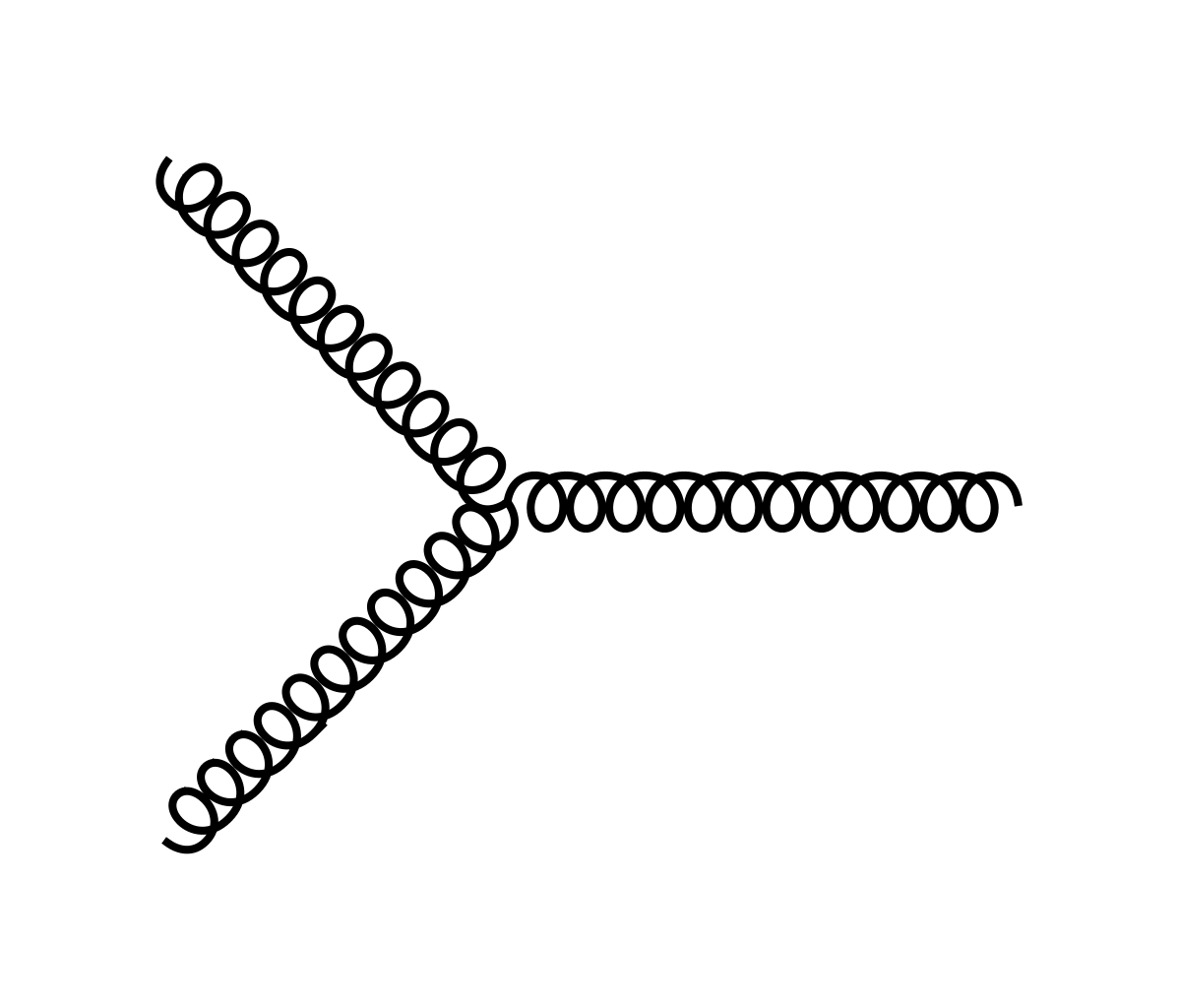} $$
\vspace{-2.1cm}
$$
\hspace{3cm}\ = \ -\ii g\, \epsilon_{\mu\nu\rho} \, F^{abc}(k,p,q)$$

\paragraph{\bf Twisted Colour-Kinematics Duality.} 
 
Like its commutative version, noncommutative Chern-Simons theory has no propagating degrees of freedom and so all gluon scattering amplitudes vanish. However, one can go slightly off-shell and consider amplitudes of `harmonic' gluon states~\cite{Szabo:2023cmv}. 
Then numerators of these amplitudes in Lorenz gauge are built from the Lie brackets of the twisted tensor product 
 $\fru(N)\otimes_{\tau}\mathfrak{vdiff}(\bbr^3)$ with the infinite-dimensional Lie algebra $\mathfrak{vdiff}(\bbr^3)$ of volume-preserving diffeomorphisms of $\bbr^3$. The generators are
\begin{equation}
 \varepsilon^a_\mu (p) \= e_p^a \ \Pi_{\mu \nu}(p)\, \partial^\nu \ ,
 \end{equation}
 where $\Pi_{\mu\nu}$ are transverse projection operators with $\Pi_{\mu \nu}(p)\, p^\mu = 0$. The commutation relations are
 \begin{equation}
 [\varepsilon^a_\mu(k) \stackrel{\star}{,} \varepsilon^b_\nu(p)] \= 
    {\tt F}_{\mu \nu\rho}^{abc}(k,p,k+p) \ \varepsilon^c_\rho(k+p) \ ,
    \end{equation} 
where the momentum dependent structure constants are given by
\begin{equation}
    {\tt F}_{\mu \nu \rho}^{abc}(k,p,k+p) \= F^{abc}(k,p,k+p) \ \Pi_{\mu \mu'}(k) \, \epsilon_{\mu' \nu' \rho} \, \Pi_{\nu' \nu}(p) \ .
\end{equation}
 
\paragraph{\bf Double Copy.}

As before, using Lie derivatives we factorize the (differential graded) Lie algebra  $\big(\Omega^\bullet(\bbr^3,\fru(N)),[-\!\ds\!-]\big)$ as the twisted tensor product 
$\fru(N)\otimes_{\tau}\Omega^\bullet(\bbr^3)$  with the (differential graded) commutative algebra $\Omega^\bullet(\bbr^3)$. This factorization is compatible with twisted colour-kinematics duality.

Replacing now the full \emph{twisted} colour-stripping by kinematic factors leads to the same double copy theory as that of ordinary Chern-Simons theory with itself~\cite{Ben-Shahar:2021zww}.
This is a non-local theory of  fields  $H_{\bar\mu\mu}=A_{\bar\mu}\otimes A_\mu$, with action functional
\begin{equation}
\widehat{S}_{{\rm CS}}[H] \= \int\, \mathrm{d}^3x \ \epsilon^{\bar \mu \bar \nu \bar\rho} \, \epsilon^{\mu \nu \rho} \, \Big(
    H_{\bar \mu \mu} \,
    \frac{1}{\square} \, \partial_{\bar \nu} \, \partial_\nu H_{\bar \rho \rho} 
    + \frac{g}{3} \, H_{\bar \mu \mu} \, H_{\bar \nu \nu} \, H_{\bar \rho \rho}
    \Big) \ .
\end{equation}

This example illustrates a general result, which in particular explains what we found in~\S\ref{sec:doublecopyNCYM}: Double copies of Moyal-deformed gauge theories coincide with the ordinary double copy theories of their commutative counterparts.

\section{Coloured Moyal Deformations of Cubic Scalar Theory}
\label{sec:Moyalscalar}

Among other things, in this section we will answer Question~1 from \S\ref{sec:doublecopyNCYM}.

\paragraph{\bf Adjoint Scalar Theory.}

Setting $\bar N=1$ in the noncommutative biadjoint scalar theory from \S\ref{sec:CKduality} leads to a non-local theory of a massless scalar field $\phi\in\Omega^0(\bbr^d,\frg)$ with rigid colour symmetry and action functional
\begin{equation}
S_{{\rm ad}}[\phi] \= \int\,\dd^dx  \ \frac12\,\phi^{a}\star\square\,\phi_{a} - \frac {\ii g}{3} \,  f^{abc} \, \phi^{a} \star \phi^{b} \star \phi^{c} \ .
\end{equation}
This theory has a standard scalar propagator and the
3-point Feynman vertex
\vspace{-3mm}
\unitlength=1.00mm
\linethickness{0.4pt}
$$ \hspace{-3.5cm}
\begin{picture}(100.00,25.00)
\thinlines
\put(50.00,12.00){\line(-1,1){6.00}}
\put(50.00,12.00){\line(-1,-1){6.00}}
\put(38.00,18.00){\makebox(0,0)[l]{{$k,a$}}}
\put(38.00,6.00){\makebox(0,0)[l]{{$p,b$}}}
\put(50.00,12.00){\line(1,0){8.00}}
\put(60.00,12.00){\makebox(0,0)[l]{$q,c$}}
\put(66.00,12.00){\makebox(0,0)[l]{{$\=-\ii g\,f^{abc}\, F(k,p,q)$}}}
\end{picture}
\vspace{-5mm}
$$

An unusual feature of the adjoint scalar theory is that it becomes free when $\theta=0$ (hence the omission of the adjective `noncommutative'). This is not what normally happens in Moyal-deformed scalar theories, where the effect of noncommutativity is to replace local interactions of an ordinary scalar field theory with non-local interactions induced by the Moyal product.

The adjoint scalar theory possesses manifest colour-kinematics duality: numerators of its amplitudes  factorize according to the untwisted tensor product $\frg\otimes\fru_\star(1)$. The kinematic algebra is the Lie algebra  $\fru_\star(1)$.

Following the relations of the twist formalism, the zeroth copy of the adjoint scalar theory replaces the Lie algebra $\fru_\star(1)=\fru(1)\otimes_\tau\Omega^0(\bbr^d)$ with $\bar\frg\otimes\Omega^0(\bbr^d)$. This results in the commutative biadjoint scalar theory from \S\ref{sec:CKduality}.

On the other hand, the double copy relation applied to the adjoint scalar theory replaces $\frg\otimes\fru_\star(1)$ with $ \fru(1)\otimes_{\bar \tau}\big(\fru(1)\otimes_\tau\Omega^0(\bbr^d)\big)$. This results in a complicated cubic theory of a scalar field $\phi\in\Omega^0(\bbr^d)$ deformed by two bivectors $\theta$ and $\bar\theta$, with action functional
\begin{eqnarray}\label{eq:DCadaction}
\widehat{S}_{{\rm ad}}[\phi] 
    \= \int\,\dd^dx  && \!  \frac12\,\phi\star \square\,\phi - \frac{\ii g} 3\, \sum_{n=0}^\infty\, \frac{(-1)^n}{(2n+1)!} \, \bar\theta^{\mu_1\nu_1} \cdots \bar\theta^{\mu_{2n+1}\nu_{2n+1}}  \\
&& \hspace{3cm} \times \ \phi\ [\partial_{\mu_1}\cdots\partial_{\mu_{2n+1}}\phi \ds \partial_{\nu_1}\cdots\partial_{\nu_{2n+1}}\phi]  \ . \nonumber
\end{eqnarray}

\paragraph{\bf Binoncommutative Biadjoint Scalar Theory.}

Taking things one step further, one can modify the double copy of the adjoint scalar theory to include colour: the scalar fields now live in the double twisted tensor product
\begin{equation}
\mathfrak{u}(N) \otimes_{\tau} \big( \mathfrak{u}( \bar N) \otimes_{\bar\tau} \Omega^0(\bbr^d) \big) \ .
\end{equation}
The interactions in this theory involve a binoncommutative product which is a binonabelian extension of the product used to construct the double copy theory for the adjoint scalar. This theory is the \emph{zeroth copy} of the noncommutative Chern-Simons theory from \S\ref{sec:CS}.

This scalar field theory is quite complicated when written down in position space in terms of an action functional, but it has a simple representation in momentum space as the standard scalar propagator with the Feynman vertex
\vspace{-3mm}
\unitlength=1.00mm
\linethickness{0.4pt}
$$ \hspace{-1cm}
\begin{picture}(100.00,25.00)
\thinlines
\put(30.00,12.00){\line(-1,1){6.00}}
\put(30.00,12.00){\line(-1,-1){6.00}}
\put(15.00,20.00){\makebox(0,0)[l]{{$k,a,\bar a$}}}
\put(16.00,4.00){\makebox(0,0)[l]{{$p,b,\bar b$}}}
\put(30.00,12.00){\line(1,0){8.00}}
\put(40.00,12.00){\makebox(0,0)[l]{$q,c,\bar c$}}
\put(48.00,12.00){\makebox(0,0)[l]{{$ \ \=-\ii g\,F^{abc}(k,p,q)\,\bar F^{\bar a\bar b\bar c}(k,p,q)$}}}
\end{picture}
\vspace{-3mm}
$$
The $n$-point colour-stripped tree-level binoncommutative biadjoint scalar amplitudes $\boldsymbol m_n^{\star\bar\star}$ can be obtained by dressing the corresponding amplitudes $\boldsymbol m_n$ of the usual biadjoint scalar theory from \S\ref{sec:CKduality} with the phase factors (\ref{eq:Thetan}) as~\cite{Szabo:2023cmv}
\begin{equation} \label{eq:bincampl}
\boldsymbol m_n^{\star\bar\star}(w|w') \= \Theta_n(w) \ \boldsymbol m_n(w|w') \ \bar\Theta_n(w') \ .
\end{equation}

\paragraph{\bf Noncommutative Yang-Mills Theory.}

We finally return to our story from \S\ref{sec:doublecopyNCYM} and reconsider the Moyal deformation of Yang-Mills theory on $\bbr^d$. It is defined by the action functional
\begin{equation}
S^\star_{{\rm YM}}[A] \= \frac{1}{2}\, \int\, \mathrm{Tr}\,\big(F_A^\star\, \wedge_\star \ast_{\textrm{\tiny $G$}}\, F_A^\star\big) \ ,
\end{equation}
where \smash{$F_A^\star = \dd A + \mbox{$\frac g2$}\,[A \stackrel{\star}{,}A]$} is the noncommutative field strength of a gauge field \smash{$A \in \Omega^1(\mathbbm{R}^{d}, \mathfrak{u}(N))$}, and \smash{$ \ast_{\textrm{\tiny $G$}}$} denotes the Hodge duality operator with respect to the open string metric $G$.

The detailed analysis of this theory is rather involved and was discussed extensively in~\cite{Szabo:2023cmv}. The double copy relations require a \emph{strictification} of the theory, which amounts to replacing the quartic interactions of gluons by cubic interactions through the introduction of suitable auxiliary fields. This is possible to do in a way that twisted colour-stripping  is compatible with colour-kinematics duality at each multiplicity order. The corresponding tree-level scattering amplitudes obey the appropriate versions of the BCJ relations. 

From (\ref{eq:KLTmomker}), (\ref{eq:KLTBfield}) and (\ref{eq:bincampl}) it follows that the $B$-field modified KLT momentum kernel $\boldsymbol S_n(w|w')_{G;\theta,\bar\theta}$ of the double copy relations for noncommutative Yang-Mills theory is sourced by the binoncommutative biadjoint scalar theory as
\begin{equation}
  \boldsymbol S_n(w|w')_{G;\theta,\bar\theta} \= \boldsymbol m^{\star\bar\star}_n(w|w')_G^{-1} \ .
\end{equation}
This answers Question~1 from \S\ref{sec:doublecopyNCYM}. The novel scalar theories introduced in this section have further interesting applications which we now describe.

\section{$\boldsymbol{\fru_\star(1)}$ Gauge Theories are Double Copy Theories}

The rank $N=1$ limits of noncommutative gauge theories are interesting interacting theories of photons. Like the adjoint scalar theory,  they become free theories in the commutative limit $\theta=0$. They have no colour symmetry, but they are invariant under the kinematic Lie algebra $\fru_\star(1)$ of noncommutative gauge transformations. These theories have long been believed to contain gravitation. Evidence is provided for instance through the interplay between gauge transformations and spacetime diffeomorphisms discussed in \S\ref{sec:CKduality}, as well as emergent gravity phenomena; see e.g.~\cite{Szabo:2006wx,Steinacker:2010rh} for reviews.

The double copy relations offer a new perspective on the emergence of gravitation: a rank one noncommutative gauge theory can be realised as the double copy of the adjoint scalar theory from \S\ref{sec:Moyalscalar} with the ordinary non-abelian version of the gauge theory in question, for any gauge algebra $\frg$. Symbolically
  \begin{equation} \label{eq:DCrank1}
\textrm{$\fru_\star(1)$ Gauge Theory} \= \textrm{Adjoint Scalar Theory} \  \otimes \ 
  \textrm{Gauge Theory} \ .
  \end{equation}
For example,  in the case of Chern-Simons theory, the left-hand side of this relation realises a theory with local symmetry under a subalgebra of spacetime diffeomorphisms, along the lines discussed in \S\ref{sec:CS}.

\paragraph{\bf Self-Dual Yang-Mills Theory and Gravity.}

The general relation (\ref{eq:DCrank1}) encompasses recent discussions of the best understood example of the double copy. Set  $d=4$  and work in coordinates $(x^+,x^-,z,\bar z)$ on $\bbr^4  \simeq  \bbr^2\times\bbc$. Turn on bivectors $\theta$ and $\bar\theta$ whose only non-zero components are equal to $\pm\, 1$ along the null $(z,x^-)$-plane. 

Then the semi-classical limit in $\bar\theta_{-z}$ of the $\bar N=1$ binoncommutative biadjoint scalar theory from \S\ref{sec:Moyalscalar} is the usual integrable deformation of self-dual Yang-Mills theory given by
the noncommutative instanton equations~\cite{Takasaki:2000vs}
\begin{equation} \label{eq:NCinst}
\square\,\phi-{\ii g}\,[\partial_z\phi\ds\partial_-\phi] \= 0 \ , 
\end{equation}
where $\phi\in\Omega^0(\bbr^4,\fru(N))$ is the Leznov prepotential representing the single polarization state of a gluon that remains after projection to the self-dual sector. 
The self-dual sector of noncommutative Yang-Mills theory has a non-trivial commutative limit to ordinary self-dual
Yang-Mills theory and arises (in Leznov gauge) from quantizing open ${\cal N}=2$
strings in a constant $B$-field background~\cite{Lechtenfeld:2000nm}. 

For $N=1$, this theory is the double copy of the adjoint scalar theory constructed in \S\ref{sec:Moyalscalar} and (\ref{eq:NCinst}) coincides with the deformed
Pleba\'nski equation for noncommutative gravity~\cite{Strachan:1992em,Takasaki:1992jf}. In the semi-classical limit in $\theta_{-z}$ it reduces to Pleba\'nski's equation $\square\,\phi =
g\,\{\partial_z\phi,\partial_-\phi\}_\theta$ for self-dual gravity, and the theory with action functional (\ref{eq:DCadaction}) is related to the closed string field theory of ${\cal N}=2$
strings~\cite{Ooguri:1991fp}. Thus self-dual $\fru_\star(1)$ Yang-Mills theory is noncommutative self-dual gravity. 

The semi-classical limit of the  adjoint scalar theory itself is self-dual Yang-Mills theory in Leznov gauge~\cite{Parkes:1992rz}, with equation of motion $\square \, \phi=g \, [\partial_z\phi,\partial_-\phi]$. It reproduces the well-known
double copy relation in the self-dual sector
\begin{equation} \label{eq:DCselfdual}
\textrm{Self-Dual Gravity} \= (\textrm{Self-Dual Yang-Mills Theory})^2 \ , 
\end{equation}
where the kinematic algebra is the infinite-dimensional Lie algebra $w_{1+\infty}$ of area-preserving diffeomorphisms of the null $(z,x^-)$-plane~\cite{Monteiro:2011pc}. Thus the adjoint scalar theory is a deformation of self-dual Yang-Mills theory
consistent with colour-kinematics duality; however, it is not integrable~\cite{Chacon:2020fmr}. On the other hand, the double copy relation (\ref{eq:DCrank1}) now reads
\begin{equation} \label{eq:DCNCselfdual}
\stackrel{\textrm{Noncommutative}}{\textrm{ Self-Dual Gravity}^{\phantom{\dag}}} {}_{\phantom{dag}}^{\boldsymbol\=} \stackrel{\textrm{Adjoint Scalar}}{\textrm{ Theory}^{\phantom{\ast}}} \ \  {}_{\phantom{dag}}^\bigotimes \ 
  \stackrel{\textrm{Self-Dual}}{\textrm{ Yang-Mills Theory}^{\phantom{\dag}}} \ {}_{\phantom{dag}}^{\boldsymbol .}
\end{equation}
In both instances (\ref{eq:DCselfdual}) and (\ref{eq:DCNCselfdual}) the integrability of gravity is ``inherited'' from the
gauge theory~\cite{Chacon:2020fmr,Monteiro:2022lwm}. 

\section{Outlook: A Braided Double Copy?}

In this contribution we have described recently developed twisted versions of colour-stripping and colour-kinematics duality that can handle the non-locality of Moyal-deformed gauge theories which induce momentum dependent structure constants in amplitudes~\cite{Szabo:2023cmv}. The result of the double copy relations is then classical, in that ordinary gravitational theories are always produced. It is natural to wonder whether there is a way in which a noncommutative theory of gravity is related to noncommutative gauge theory. We assert that the answer is affirmative, through a suitable variant of the double copy.

Moyal-deformed theories of gravity have been developed with ``twisted'' diffeomormphism symmetry, though it is unclear whether these models can arise in string theory; see e.g.~\cite{Szabo:2006wx} for a review. These theories lie in the broad class of `braided quantum field theories', which are theories with braided symmetries that can be defined and treated in the new language of \emph{braided homotopy algebras}~\cite{DimitrijevicCiric:2021jea,Giotopoulos:2021ieg}. The key point is that if one works with {braided homotopy algebras}, then \emph{untwisted} colour-stripping and factorization are possible, with the kinematic factors now retaining information about noncommutativity in the form of (differential graded) braided commutative algebras. 

Hence it should be possible to realise twisted noncommutative gravity theories through
\begin{itemize}
\item[] \smash{\sf \underline{Conjecture:}} \ There is a braided version of the double copy construction in terms of factorizations into braided kinematic algebras and braided noncommutative field theories such that
\begin{equation}
\textrm{Twisted Noncommutative Gravity} \= (\textrm{Braided Gauge Theory})^2 \ .
\end{equation}
\end{itemize}

\paragraph{Acknowledgements.}

The author thanks Guillaume Trojani for the collaboration on~\cite{Szabo:2023cmv}. This article is based upon work from COST Actions CaLISTA CA21109 and THEORY-CHALLENGES CA22113 supported by COST (European Cooperation in Science and Technology).

%
%

\end{document}